\pdfoutput=1
\documentclass[prb,onecolumn]{revtex4-1}
\usepackage{graphicx}

\normalfont
\usepackage{amsmath}
\usepackage{upgreek}% bold math
\usepackage{epstopdf}
\newcommand{\micro}{${\upmu}$}
\newcommand{\micron}{${\upmu}$m }
\newcommand{\vect}[1]{\boldsymbol{#1}}

\begin{document}

%\preprint{APS/123-QED}

\title{Two-dimensional metal-chalcogenide films in tunable optical microcavities}% Force line breaks with \\

\author{S.Schwarz$^1$}
\email{s.schwarz@sheffield.ac.uk}
\author{S. Dufferwiel$^1$}
\author{P. M. Walker$^1$}
\email{p.m.walker@sheffield.ac.uk}
\author{F. Withers$^2$}
\author{A. A. P. Trichet$^3$}
\author{M. Sich$^1$}
\author{F. Li$^1$}
\author{E. A. Chekhovich$^1$}
\author{D. N. Borisenko$^4$}
\author{N. N. Kolesnikov$^4$}
\author{K. S. Novoselov$^2$}
\author{M. S. Skolnick$^1$}
\author{J. M. Smith$^3$}
\author{D. N. Krizhanovskii$^1$}
\author{A. I. Tartakovskii$^1$}
\email{a.tartakovskii@sheffield.ac.uk}

\affiliation{$^1$Department of Physics and Astronomy, University of Sheffield, Sheffield S3 7RH, UK}

\affiliation{$^2$School of Physics and Astronomy, University of Manchester, Manchester M13 9PL, UK}

\affiliation{$^3$Department of Materials, University of Oxford, Parks Road, Oxford OX1 3PH, UK}

\affiliation{$^4$Institute of Solid State Physics, Russian Academy of Sciences, Chernogolovka 142432, Russia}

\date{\today}

\begin{abstract}
\textbf{Quasi-two-dimensional (2D) films of layered metal-chalcogenides have attractive optoelectronic properties. However, photonic applications of thin films may be limited owing to weak light absorption and surface effects leading to reduced quantum yield. Integration of 2D films in optical microcavities will permit these limitations to be overcome owing to modified light coupling with the films. Here we present tunable microcavities with embedded monolayer MoS$_2$ or few monolayer GaSe films. We observe significant modification of spectral and temporal properties of photoluminescence (PL): PL is emitted in spectrally narrow and wavelength-tunable cavity modes with quality factors up to 7400;  PL life-time shortening by a factor of 10 is achieved, a consequence of Purcell enhancement of the spontaneous emission rate. This work has potential to pave the way to microcavity-enhanced light-emitting devices based on layered 2D materials and their heterostructures, and also opens possibilities for cavity QED in a new material system of van der Waals crystals.}
\end{abstract}

\maketitle

%\tableofcontents

\section{Introduction}

The discovery and research into remarkable properties of two-dimensional (2D) sheets of carbon \cite{NovoselovScience2004}, graphene, has sparked interest in other 2D materials such as metal chalcogenides (MCs) \cite{NovoselovPNAS2005,WangNatNano2012,XuNatPhys2014}. MCs are a large family of compounds that include layered van der Waals crystals. Among them are many semiconducting materials with a variety of band-gap energies attractive for use in opto-electronics and also suitable for fabrication of 2D films with thicknesses down to a single unit cell \cite{NovoselovPNAS2005,WangNatNano2012,XuNatPhys2014,BritnellScience2013,GeimNature2013,PospischilNatNano2014}. Recent progress in device fabrication resulted in thin-film transistors \cite{RadisavljevicNatNano2011}, light emitting devices and photodetectors utilizing novel types of heterostructures\cite{BritnellScience2013,GeimNature2013,PospischilNatNano2014} compatible with a wide variety of materials including dielectric, polymer and flexible substrates. 

For light-emitting applications, some of the layered materials such as for example molybdenum and tungsten dichalcogenides need to be thinned down to a single atomic unit cell (e.g. S-Mo-S for MoS$_2$), as only in this form do they behave as direct band-gap semiconductors \cite{MakPRL2010,SplendianiNanoLett2010,RossNatComm2013}. Layered III-VI crystals such as GaSe, InSe etc are also attractive for 2D photonics as they exhibit direct band-gaps in films with a variety of thicknesses from a few to tens of nanometers \cite{HuACSNano2012,MuddAdvMat2013}, thus enabling flexibility in the device design. However, for materials of both types, use of thin films in photonic applications will require overcoming relatively low emission and absorption efficiency generally characteristic of thin layers of any semiconducting material.

In traditional direct-band semiconductors, practical ways to improve the device efficiency include (i) embedding thin films in thicker heterostructures [such as e.g. in GaAs/AlGaAs quantum wells (QWs)]\cite{HolonyakIEEE1980}; (ii) using structures with multiple QWs \cite{Nakamura1996}; (iii) enhancing light absorption/emission in a semiconductor by embedding it in a microcavity (Fabry-Perot, photonic crystal etc)\cite{Koyama2006,HennessyNature2007,Nowak2014} or a waveguide\cite{HolonyakIEEE1980}. All methods (i) to (iii) are usually combined in light-emitting diodes and lasers\cite{Koyama2006,HolonyakIEEE1980,Nakamura1996}. In respect to novel layered materials, fabrication of heterostructures required for approaches (i) and (ii) is still in the early stages of development \cite{BritnellScience2013,GeimNature2013,PospischilNatNano2014}. So far, approach (iii) has mainly been applied to graphene\cite{FurchiNanoLett2012,EngelNatureComm2012}, showing significant enhancement of light absorption when placed in photonic structures. Enhancement of PL in semiconducting MoS$_2$ and WSe$_2$ films coupled to photonic crystal cavities has also been demonstrated \cite{GanAPL2013,Wu2014}. More recently, strong exciton-photon coupling has been observed in a dielectric cavity containing synthesized MoS$_2$ layers, emphasizing the dramatic effect optical cavities may have on the properties of thin films \cite{LiuArxiv2014}.

Here we demonstrate tunable external cavity devices where thin films are deposited on a flat dielectric distributed Bragg reflector (DBR) and the cavity is completed by another DBR having a concave shape \citep{Dolan2010}. We show that in this mirror configuration it is possible to form cavity modes confined in three dimensions and having mode volumes as low as 1.6$\mu$m$^3$ and quality factors, $Q$, up to 7400. This design allows tunability of the cavity size by adjusting the vertical displacement of the two mirrors and, therefore, allows spectral matching of the cavity mode wavelength with that of the emitter embedded in the device: we demonstrate wavelength tuning over 80 nm limited by the spectral width of the emitter only. We realize such devices for single atomic layers of molybdenum disulphide (MoS$_2$) and few monolayer thick gallium selenide (GaSe) films and observe major modification of the emission spectra and strong peak PL intensity increase compared with the same films but with the top mirror moved out of the optical path. Using time-resolved PL spectroscopy, we directly observe the Purcell enhancement of the spontaneous emission rate for films placed in the cavity by measuring the shortening of PL life-times by a factor of 10 in GaSe films. The shortening of the radiative life-time is further confirmed in continuous-wave power-dependence PL measurements: we observe clear PL saturation at high powers for a GaSe film on a flat DBR, and no saturation for PL of the same film inside a cavity. By performing finite-difference time-domain (FDTD) calculations, we show that the observed PL intensity enhancement is also due to the highly directional cavity mode emission having narrow angular distribution and enabling improved light coupling to the collection optics.

\begin{figure*}
\includegraphics[width=0.8\textwidth]{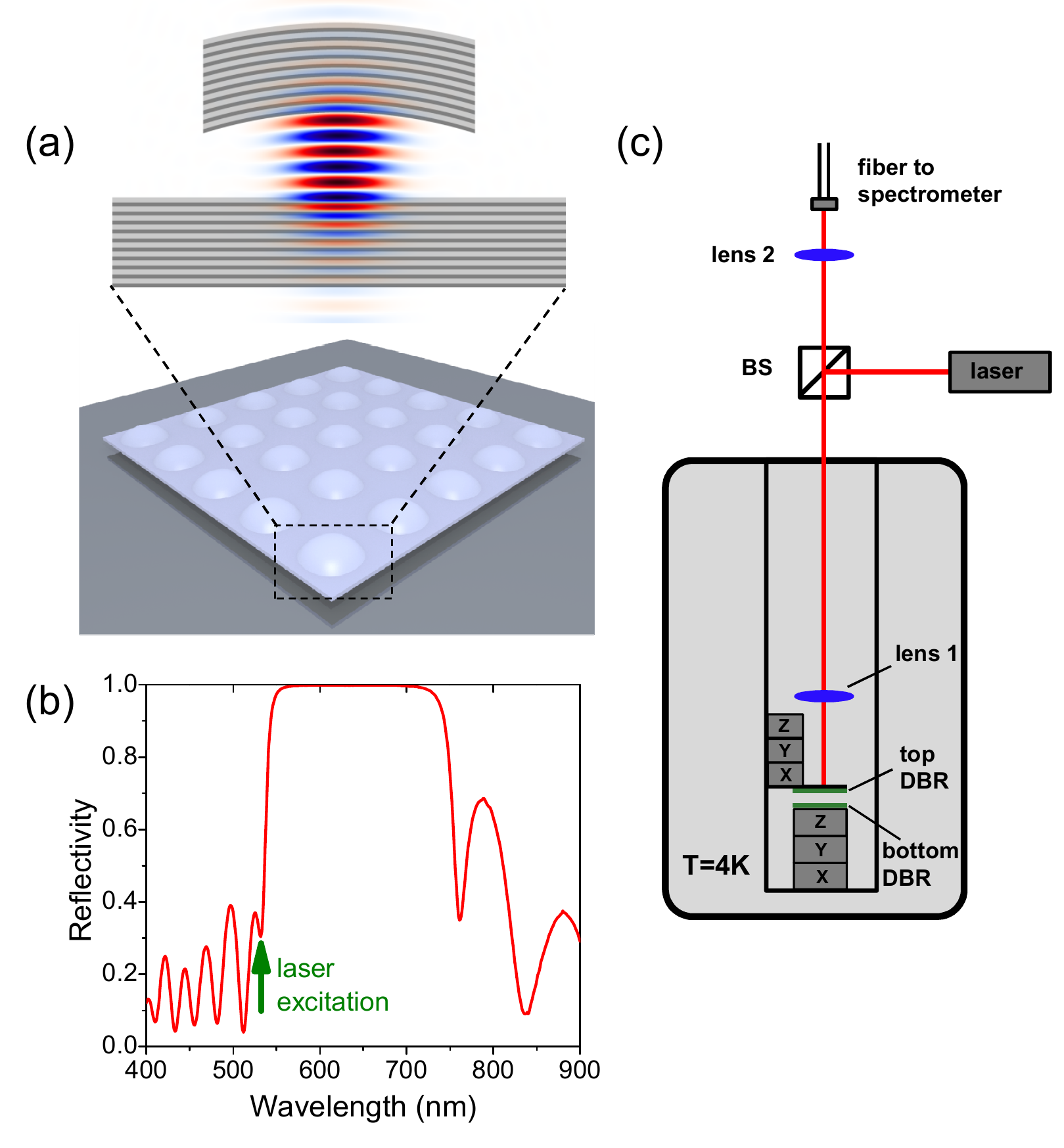}
   \caption{\label{Fig1} {\bf Microcavity design and schematic diagram of the experimental setup.} (a) The microcavity is formed by a planar and a concave dielectric mirror having precisely controllable vertical displacement. Monolayer MoS$_2$ and thin sheets of GaSe are transferred onto the flat mirror. The concave mirror provides the cavity mode confinement is three dimensions as shown in the calculated electric field distribution. (b) The reflectivity of the deposited DBR shows a wide stop-band in a range of $\approx$ 550-720 nm. The cw 532 nm laser excitation used for most of the measurements is shown with a vertical arrow and is spectrally just outside the stop-band. Transmission of 70\% is observed for 532 nm. (c) Experimental setup showing the tunable open-access microcavity placed at $T=4.2$K inside a liquid helium cryostat.}
\end{figure*}

Fig.\ref{Fig1}(a) schematically shows our microcavity system formed by a planar bottom DBR and a top DBR having a concave shape\cite{DufferwielAPL2014}. Both mirrors have reflectivity exceeding 99\% for the design wavelength of 650 nm, and a stop band of 200 nm as seen from the normal incidence reflectivity spectrum in Fig.\ref{Fig1}(b). The electric field distribution inside the cavity in Fig.\ref{Fig1} is calculated using an FDTD method, and shows confinement of the mode in both vertical and lateral directions \cite{SI}. In our experiments we achieve confined cavity mode volumes down to 1.6$\mu$m$^3$. Arrays of concave mirrors with different nominal radii of curvature, $R_c$, from 5.6 to 25 $\mu$m are patterned in $\approx$ 5 mm $\times$5 mm silica substrates as schematically shown in Fig.\ref{Fig1}, and concave SiO$_2$/TiO$_2$ DBRs are fabricated by coating the substrates with alternating dielectric layers \cite{SI}. 

In our set-up both the bottom flat DBR and the substrates with arrays of concave DBRs are mounted on individual XYZ nano-positioners (see Methods and Ref.\cite{SI} for more details). This enables accurate alignment of individual concave mirrors with any given area of the 2D film deposited on the bottom DBR. Importantly, this also permits control of the vertical cavity length, $L_{cav}$, enabling wavelength tuning of the cavity modes as described with the following expression: 
\begin{equation}
\frac{1}{\lambda_{q,m,n}} = \frac{1}{2L_{opt}}\left[q+\frac{m+n+1}{\pi}\arccos\left(1-\frac{L_{cav}}{R_c}\right)\right]
\label{Eq1}
\end{equation}
Here, $q$, $m$ and $n$ are longitudinal ($q$) and transverse ($m$, $n$) mode numbers for Hermite Gauss cavities. $L_{opt}$ is the effective optical length of the cavity, which differs from the physical length $L_{cav}$ due to penetration of the cavity mode into the DBRs. In our case, $L_{opt}=L_{cav}+1.25\mu$m, which is determined experimentally so that the mode wavelengths can be described with Eq.\ref{Eq1}. 

The investigated thin layers of MoS$_2$ and GaSe are produced by mechanical cleaving of bulk crystals. All PL results were obtained at low temperature $T$=4.2 K. The measurements were carried out using a low-T micro-PL set-up [see the diagram in Fig.\ref{Fig1}(c)] with a typical excitation laser spot on the sample surface of $\approx$7 $\mu$m. The whole external cavity set-up including the nano-positioners is cooled down to $T$=4.2 K in a He gas-exchange cryostat [Fig.\ref{Fig1}(c)]. The PL from the 2D films is collected from the top side of the microcavity (the side of the concave DBR) through a lens with a diameter 5 mm and then focused outside the cryostat on to an optical fiber leading to a spectrometer and a charge-coupled device.\\

\section{Results}

\subsection{Spectral modification of PL by the cavity}

\begin{figure*}
\includegraphics[width=0.7\textwidth]{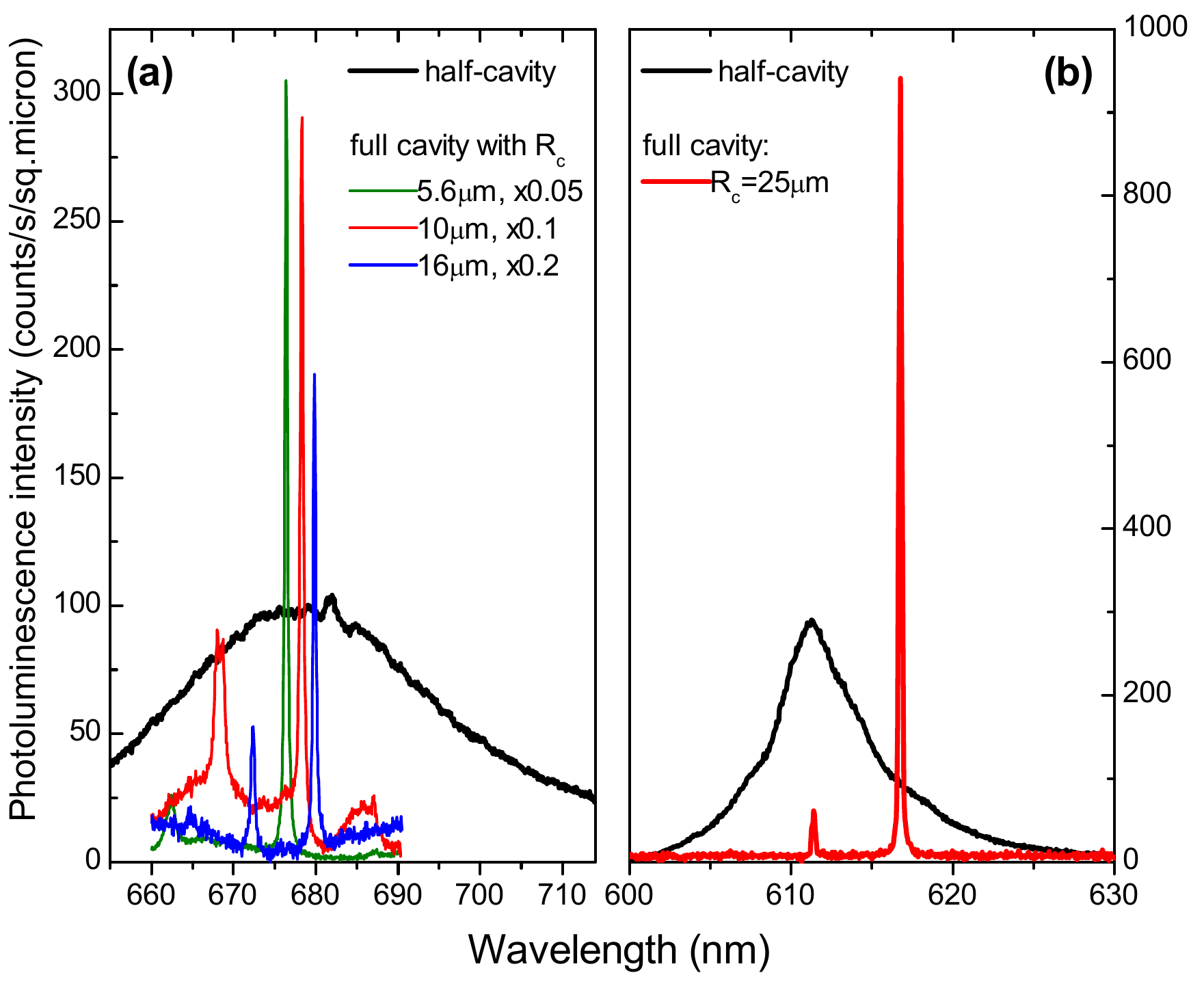}
\caption{\label{Fig2} {\bf Thin film and cavity mode photoluminescence spectra.} The broadband photoluminescence (PL) is collected from 2D films placed on the flat bottom DBR with the concave DBR moved out of the optical path. The narrow cavity modes are measured in PL when the microcavity is formed by placing the concave top mirror in the optical path. PL is measured at $T=4.2$K. (a) A PL spectrum of a monolayer MoS$_2$ (black) is shown together with the cavity emission for the top mirror radii of curvature ($R_c$) of 5.6 $\mu$m, 10 $\mu$m and 16 $\mu$m. (b) PL of GaSe 43 nm film and cavity PL in resonance with the low energy shoulder of the GaSe spectrum. Both in (a) and (b), the PL intensity is presented in counts per second per square micron of the emitter (see text for explanation of normalization procedure). 
}
\end{figure*}

Fig.\ref{Fig2} shows low-T PL spectra for MoS$_2$ and GaSe films excited by a 532 nm cw laser. Broad spectra are obtained when the films are placed on the bottom DBR with the top DBR moved out of the optical path (the 'half-cavity' configuration). For this measurement, an excitation power density on the film of 6.5 kW/cm$^2$ was used. The PL linewidth for both materials is determined by inhomogeneous broadening caused by variations of the exciton potential within the film area illuminated by the laser. In Fig.\ref{Fig2}(a) PL from a MoS$_2$ monolayer is presented stretching over a wide range from 640 to 740 nm (1.67-1.94 eV) \cite{MakPRL2010,SercombeSciRep2013}. Fig.\ref{Fig2}(b) shows PL for a 43 nm thick GaSe film with a spectrum in the range 600-625 nm (1.98-2.07 eV) corresponding to the emission from localized exciton states, which probably occur due to the inter-layer stacking defects as was reported previously\cite{CapozziPRB1986}. 

The cavity is formed by moving the top mirror above the optically excited area of the film, with the laser coupled through the top DBR (the 'full-cavity' configuration). In this configuration the PL emitted by the 2D films is coupled into cavity modes having typical Q-factors of a few 10$^3$ and for some mirrors reaching up to 7400 (see Supplementary Information, Ref.\cite{SI}). Although the laser spot is similar to the experiments with the half-cavities, the waist diameter of the longitudinal modes is $\approx$ 1 $\mu$m \cite{DufferwielAPL2014}. Thus in the case of the full cavity, PL is collected from a film area about 50 times smaller than in the case of a film placed on a planar DBR. This is accounted for in Fig.\ref{Fig2} by using units of counts per second per square micron for PL intensity: the emission is assumed to originate from a spot with diameter of 1 $\mu$m and 7 $\mu$m for the 'full-' and 'half-cavity' configurations, respectively (see below for a more precise calculation of the waist diameter using FDTD). 

Fig.\ref{Fig2} shows several examples of spectra observed for 2D film PL emission into the longitudinal cavity modes. PL is measured for the same laser excitation density of 6.5 kW/cm$^2$ incident on the film (this value is obtained taking into account that there is about 70\% transmission through the top DBR at 532 nm). The spectra containing sharp peaks in Fig.\ref{Fig2}(a) show PL measured for a cavity with a MoS$_2$ monolayer. The strongest peaks around 675-680 nm correspond to longitudinal modes described in Eq.\ref{Eq1} with $m,n$=0. A few weaker PL peaks corresponding to modes with $m,n\neq$0 are also observed at shorter wavelength. The green, red and blue lines show spectra measured for different top mirrors with nominal radii of curvature, $R_c$, of 16, 10 and 5.6 $\mu$m, respectively. The measured $Q$-factors for the strongest modes in the spectra are 4000, 3000 and 1800, respectively.

For the full cavity having a concave DBR with $R_c$=16, 10 and 5.6 $\mu$m, the intensity (in counts/s/$\mu$m$^2$) at the wavelengths corresponding to the mode PL peaks is greater than the MoS$_2$ monolayer PL by $\approx$ 10, 30 and 60 times, respectively. Note, that for each experiment one of the longitudinal modes was tuned in resonance with MoS$_2$ PL roughly at the same wavelength. For this, the vertical size of the cavity was adjusted to values varying for different $R_c$: PL measurements in Fig.\ref{Fig2}(a) were taken for $L_{cav}\approx$ 1.9 $\mu$m for the mirrors with $R_c$=5.6 and 10 $\mu$m, and $L_{cav}\approx$ 2.9 $\mu$m for the mirror with $R_c$=16 $\mu$m. PL enhancement by a factor of 10 is also observed for the GaSe film in Fig.\ref{Fig2}(b). Here a cavity with $L_{cav}\approx$ 4.2 $\mu$m and a concave mirror with $R_c$=25 $\mu$m was used, and a $Q$-factor of 3100 was observed.

%{\bf Wide-range tuning of the mode wavelength}

By changing the distance between the two mirrors, and therefore the cavity length, the cavity resonance wavelength, $\lambda_{q,m,n}$, is tuned according to Eq.\ref{Eq1}. In experiment, this is achieved by changing the voltage on the piezo element of the nano-positioner where the top DBR is attached. Fig.\ref{Fig3} demonstrates the tuning for a cavity with a MoS$_2$ monolayer. It is seen that the cavity modes can be tuned over a range of about 40 nm by gradually changing the cavity length. The bright peak marked in the plot corresponds to a fundamental longitudinal mode with $q$=12 for the optical cavity length of around $L_{cav}$=4.2$\mu$m. Apart from this bright longitudinal mode, several other transverse modes with $m,n \neq$0 are also visible (see the bottom left of the graph). The emission of the MoS$_2$ monolayer is centered at around 680 nm and the intensity of the cavity modes is enhanced when they are tuned past this spectral window. 
Using a similar pair of mirrors we were also able to tune the modes around the emission window of GaSe down to $\approx$600 nm. In general, the tuning is only limited by the size of the stop band, which is $\approx$200 nm in our case [see Fig.\ref{Fig1}(b)].

\subsection{Temporal modification of PL by a cavity: Purcell enhancement}

The PL enhancement observed in Fig.\ref{Fig2} can arise as a combined result of angular redistribution of light emission and enhancement of the spontaneous emission rate (Purcell enhancement). Direct evidence for the Purcell enhancement can be obtained by comparing the radiative lifetime for the emitter with and without the effect of the cavity. This information may be gained from time-resolved PL measurements, which we carry out for the GaSe thin films. Fig.\ref{Fig4}(a) shows PL decay curves for a 43 nm GaSe film excited with a pulsed laser tuned to 415 nm: the blue curve corresponds to a half-cavity and red to a full-cavity configuration, respectively. In both configurations the time-averaged excitation density was 2.6kW/cm$^2$. For the full cavity we use $L_{cav}\approx$2.9$\mu$m and $R_c=$10$\mu$m. The cavity length is adjusted in such a way that a longitudinal cavity mode is coupled into the low energy shoulder of the GaSe emission at around 613 nm similarly to the case depicted in Fig.\ref{Fig2}(b). 

As seen in Fig.\ref{Fig4}(a), the blue curve exhibits decay with a life-time of $\tau_{hc}\approx$700 ps, whereas in the full-cavity configuration $\tau_{fc}\approx$70 ps is observed (both times are obtained using fitting with mono-exponential decay functions). This measurement signifies the enhancement of the spontaneous emission rate in the tunable cavity device. 

Fig.\ref{Fig4}(b) presents further evidence for the enhancement of the radiative recombination rate in a microcavity. Here we show PL power-dependences of the 43 nm thick GaSe film in the half- and full-cavity configurations measured with a 532 nm cw laser. For the full cavity we use the same $L_{cav}$ and $R_c$ as for the time-resolved experiments. The plot shows integrated PL intensity calculated under the full-cavity mode spectrum centered at 613 nm and having a FWHM of 0.15 nm (black squares), and under the same bandwidth of GaSe PL at the same center wavelength in the half-cavity case (blue squares). The reduced transmission of the green laser through the top DBR is taken into account, and the data are plotted as functions of the actual laser power incident on the GaSe film. As in previous figures, the PL intensity normalized by the area of the spot on the film that emits light is adopted here. In the case of the half-cavity configuration, a clear sub-linear PL intensity growth with power is seen. PL saturation is observed at around $P$=10 mW, indicating that the optical pumping rate exceeds the relatively low recombination rate $1/\tau_{hc}$ ($\tau_{hc}\approx$700 ps) of the localized exciton states in GaSe \cite{CapozziPRB1986,TaylorJofPC1987}. This saturation effect is similar to the state-filling phenomenon in semiconductor quantum dots \cite{RaymondPRB1999}.

In contrast to this, a much stronger PL increase with power is found for the cavity mode in the full-cavity configuration. No PL saturation is observed, indicating that the spontaneous recombination rate remains higher than the excitation rate in the whole range of powers used. This striking difference clearly indicates that the carrier radiative life-time is markedly shorter in the case when the full cavity is formed \cite{HappPRB2002}, the effect quantified with the 10-fold PL decay time shortening in Fig.\ref{Fig4}(a).

\subsection{Calculation of the geometrical PL enhancement and Purcell factors in a cavity}

The observed enhancement of PL intensity in a cavity may be due to a combination of the enhanced spontaneous emission rate and increased directionality of the cavity mode towards the collection optics. To estimate the contribution of the mode directionality, the angular width of the fundamental cavity mode was calculated and compared with the radiation pattern for a dipole placed directly on the lower DBR but with no upper mirror using the finite difference time domain (FDTD) technique~\cite{meep} (see Methods and Ref.\cite{SI} for more details).  

Figure~\ref{Fig5} shows the simulated power radiated per unit solid angle in the far field for several mirror radii-of-curvature and separations corresponding to the experimental conditions as well as for the case without a top mirror. Only the upward flux is shown although the total flux radiated in all directions was calculated. We consider three cavities with parameters corresponding to the devices experimentally measured in Fig.\ref{Fig2}(a): cavities C1, C2 and C3 are used having $R_c$ of 16, 10 and 5.6 $\mu$m and  $L_{cav}$ of 2.98, 2.05 and 1.82 $\mu$m, respectively.

In Fig.~\ref{Fig5} it is observed that in the case of no top mirror present, the PL is emitted in a broad range of angles, whereas markedly more directional distributions are observed for the three cavities. We estimate that for the light to be collected by the objective placed above the microcavity [lens L1 in Fig.\ref{Fig1}(c)], light should be emitted within a 16.7$^o$ cone. We then obtain that in the case where no top mirror is present only 9\% of the total power is collected through the objective. We find that for the cavities C1, C2 and C3, the fractions of power within $\pm$16.7 degrees are 47\%, 44\% and 20\% respectively. The marked lowering for the cavity with the smallest $R_c$ occurs because of scattering at the discontinuity between curved and planar regions of the top mirror. Note, that in the real cavity structures this effect may not be as pronounced, as this discontinuity may be less sharp than in the model. We thus find, that the directionality of the the modes of cavities C1, C2 and C3 will lead to 'geometrical' enhancements factors, $F_g$ of 5.3, 4.9 and 2.2 respectively in the observed intensity compared to the case with no top mirror present. 

Further to the angular distribution, the FDTD method employed above can be used to calculate the volumes, $V$ and $Q$-factors for the optical modes in the microcavity. This allows calculation of the Purcell enhancement factors, $F_P$ (see Methods). Using the calculated values for $V$ and $Q$, we obtain the following Purcell factors for the three considered cavities: $F_{P1}$=41, $F_{P2}$=59 and $F_{P3}$=70. We also calculate the Purcell factor $F_P$=51 for the cavity used 
in the time-resolved measurements with $L_{cav}$$\approx$2.9$\mu$m and $R_c$=10$\mu$m. 

We note that these values for $F_P$ are for the emitter in perfect spatial and spectral overlap with the cavity mode. Thus these values provide an upper limit to the Purcell factor which may be achieved in a cavity. In the experiment, 
we deal with an inhomogeneously broadened ensemble of spectrally narrow emitters distributed in a few $\mu$m diameter area of a 2D film. This means that spatial and spectral averaging of the Purcell enhancement will occur and a value of $F_P$ lower than given above is expected (see Ref.\cite{GerardPRL1998} and Methods) as discussed in the next subsection.

\section{Discussion}

First we note that the effect of non-radiative processes needs to be taken into account for the interpretation of the experimentally measured data and comparison with the theory. Using a simple approximation, we will take into account only three processes: optical pumping, a non-radiative decay with a time $\tau_{nr}$ and a radiative decay with $\tau_r/F_P$ ($F_P$=1 for a bare thin film and $F_P>$1 is expected for emitters in resonance with the cavity mode). The 'state-filling' effects as observed in Fig.\ref{Fig4}(b) will be neglected. It is easy to show that the cw PL intensity $I_{PL}\propto \tau_{nr}/(\tau_{nr}+\tau_r/F_P)$ and the characteristic PL decay time $\tau_{PL}= \tau_{nr}(\tau_r/F_P)/(\tau_{nr}+\tau_r/F_P)$. Note, that if $\tau_{r}/F_P \gg \tau_{nr}$, $\tau_{PL}\approx\tau_{nr}$ and the Purcell enhancement will not be observable in time-resolved PL. 

In our experiments, direct evidence for the Purcell enhancement is obtained for the GaSe film in Fig.\ref{Fig4}, where a marked shortening of $\tau_{PL}$ is observed. It can be shown that the ratio of $\approx$10 of the experimentally measured $\tau_{PL}$ values in the half- and full-cavity configurations corresponds to the lower limit for the Purcell enhancement factor.  This limit is reached if $\tau_{nr} \gg \tau_{r}$. The observation of the PL life-time shortening implies that at least $\tau_{r} \alt \tau_{nr}$. Further indirect evidence that this condition is well satisfied for the localized excitons in GaSe measured in our experiments, is that in some thin films we also observe PL from extremely long-lived defect states in the same wavelength range with PL life-times of a few micro-seconds, as was also reported previously \cite{TaylorJofPC1987}.   

\begin{figure*}
\includegraphics[width=0.55\textwidth]{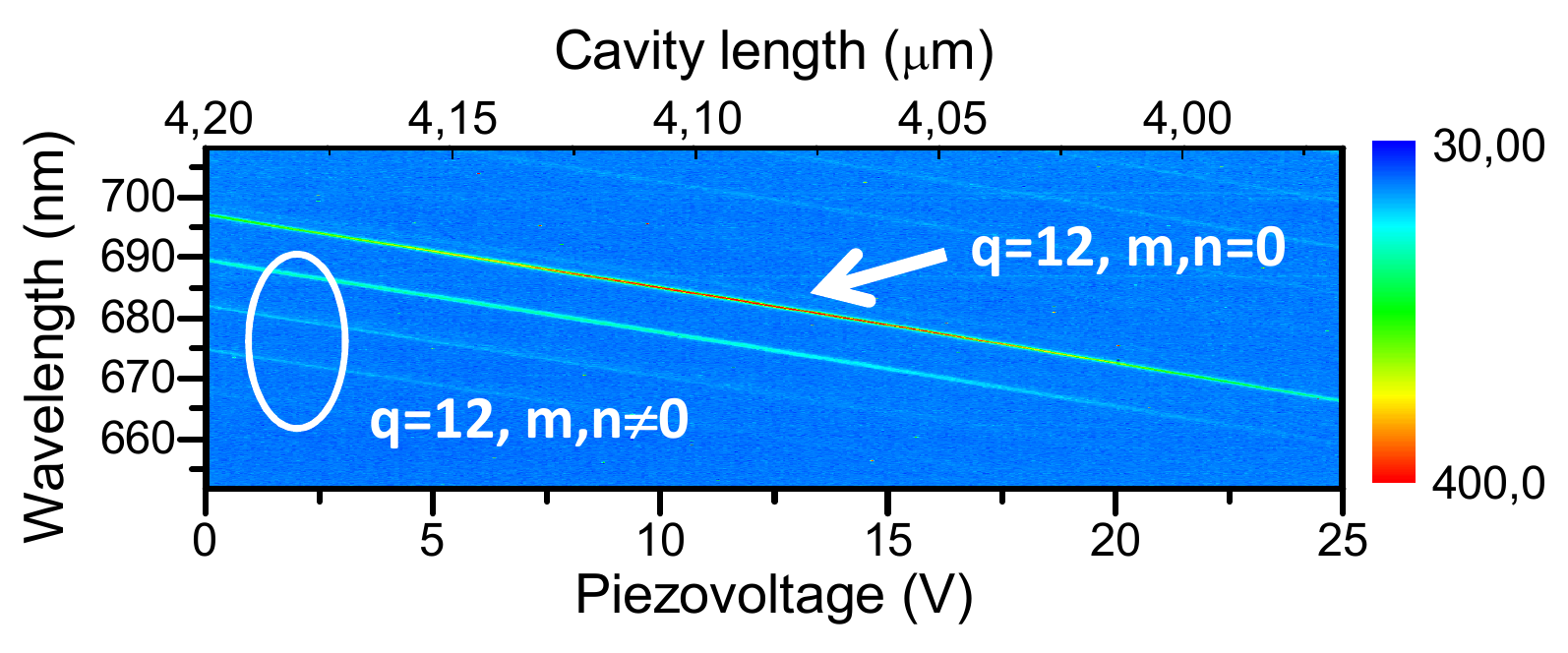}
\caption{\label{Fig3} {\bf Tuning of the mode wavelength by adjusting the vertical length of the cavity.} The distance between the DBRs is adjusted by changing the applied voltage on the piezo-nanopositioner. The figure shows PL map obtained for such length tuning for the cavity containing a monolayer MoS$_2$ film.  PL of the modes is observed in a wide spectral range overlapping with the PL of the MoS$_2$ film:  the mode PL is enhanced when in resonance with MoS$_2$ emission, showing weak coupling between the emitter and the cavity.}
\end{figure*}

\begin{figure*}
\includegraphics[width=0.8\textwidth]{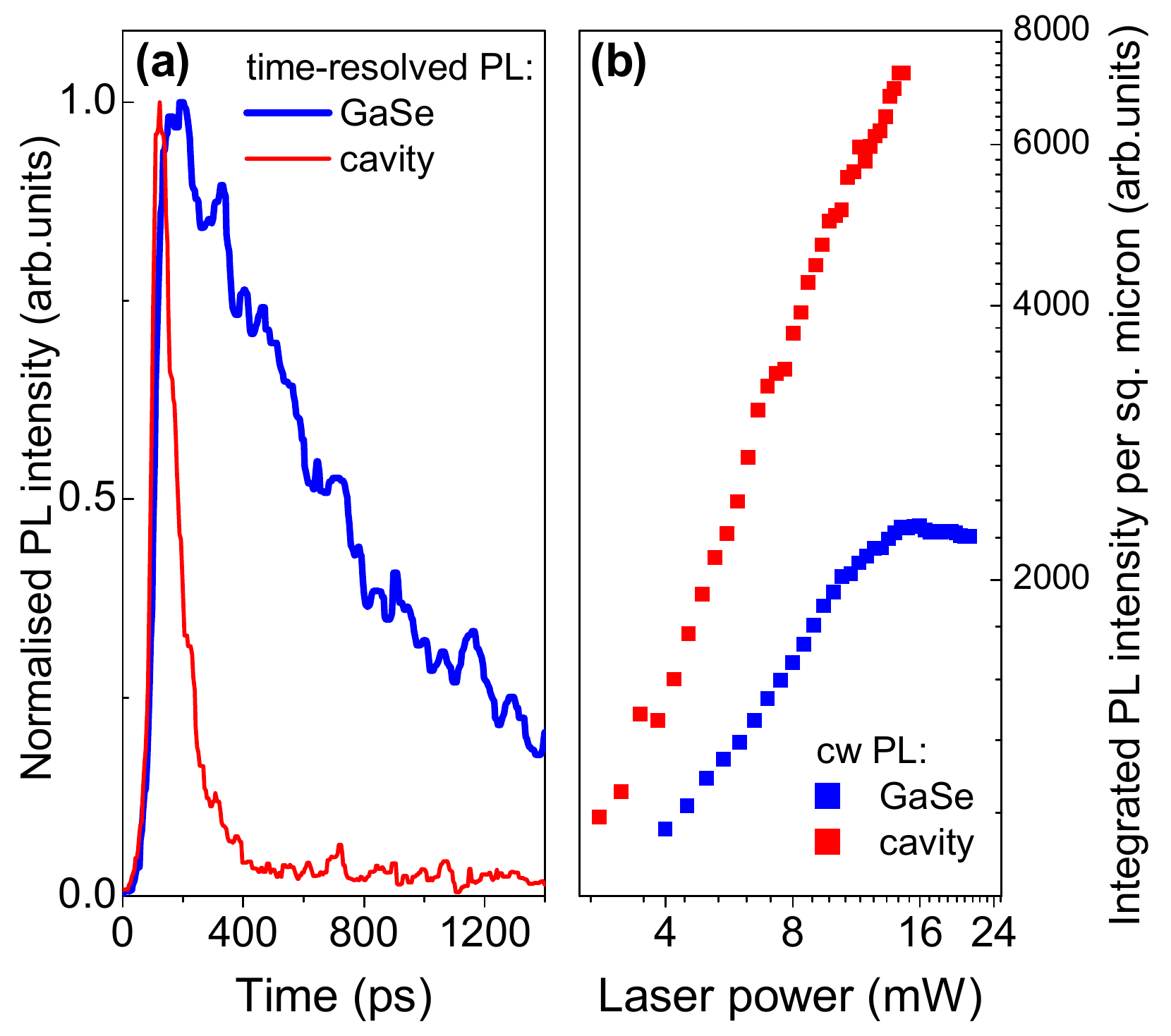}
\caption{\label{Fig4} {\bf Observation of Purcell enhancement for GaSe thin film emission in a tunable microcavity.} (a) PL traces measured at 613 nm using the streak-camera. PL is measured at $T=4.2$K using excitation with a pulsed laser at 415 nm. The blue trace shows GaSe PL decay in a film in the half-cavity configuration with a lifetime of 700 ps. The red curve is for the cavity PL decay, for which a lifetime of 70 ps is extracted. (b) Photoluminescence power dependence for a GaSe film in a microcavity. PL measurements were performed at $T=4.2$K by varying the excitation power of a continuous-wave 532 nm laser. At high laser powers, clear saturation of the GaSe film PL is observed (blue squares) when measured without the top concave mirror. In the full cavity, the cavity mode PL (red squares), fed by the GaSe film emission, shows no saturation. In the graph, the excitation power of the cavity mode is corrected by the measured transmission through the top mirror.}
\end{figure*}

In the case of MoS$_2$ monolayer films $\tau_r/F_P \gg \tau_{nr}$ \cite{Xu2014,MakPRL2010} for reasonable values of $F_P$. In this case, non-radiative processes dominate and the Purcell enhancement is not expected to be observable in time-resolved PL measurements. On the other hand, it is easy to show that in the absence of 'state-filling' effects (as is the case for MoS$_2$ \cite{SercombeSciRep2013}) and for $\tau_r/F_P \gg \tau_{nr}$, the cw PL intensity $I_{PL}\propto F_P$ in a cavity. Taking into account the geometrical enhancement due to the more directional cavity PL emission into the collection optics, $I_{PL}\propto F_P F_g$. This can be used to show that Purcell enhancement is present in all of the cavities with the MoS$_2$ monolayer film shown in Fig.\ref{Fig2}(a). As shown above, the  'geometrical' enhancement factors $F_g$ for idealized cavities with parameters corresponding to our experiments lie between 2.2 and 5.3.  By dividing the experimentally observed total PL enhancements of 10, 30 and 60, reported in Fig.\ref{Fig3}, by the maximum expected $F_g$ of $\approx$5, we obtain the lower limit for the PL enhancements caused purely by the Purcell effect. Thus lower limits of the experimental values for $F_P$ from $\approx$2 to $\approx$12 are estimated.

For cavities investigated in both the time-resolved and cw measurements, the calculated maximum Purcell enhancements in the range of 41 to 70, as found in the previous subsection, exceed those observed in the experiment. One of the reasons for this discrepancy is the lowering of the measured $Q$-factors from their calculated values occurring as a result of cavity imperfections, which is a typical observation in various microcavity designs. In our work the calculated $Q$-factors are $\approx$2.5 times larger than the measured values.  

Furthermore, in order to reflect the finite in-plane size of the mode and varied coupling to the mode of independent emitters at different wavelengths, spatially and spectrally averaged Purcell enhancement should be calculated\cite{GerardPRL1998}. In general the averaged Purcell factors are lower than the maximum values for the perfect spatial and spectral emitter-mode overlap. The averaging procedure requires summation of $I_{PL}$ (in the case of cw PL enhancement) or of the simulated PL decay curves (in the case of lifetime shortening) over the emitters in the ensemble. The total PL or average characteristic PL decay time may then be determined. However, the result is sensitive to the ratio $(\tau_r/F_P)/\tau_{nr}$ since both $\tau_{PL}$ and $I_{PL}$ depend on these parameters. At this stage, insufficient information exists about the non-radiative decay processes in 2D films for such averaging to be carried out accurately.

\section{Conclusions}

In conclusion, we report on fabrication of tunable dielectric microcavities confining the photonic field in all three directions and comprising 2D films of metal-chalcogenides, and show that the films' spectral and temporal PL properties are strongly modified by the cavity. We observe strong cw PL enhancement for MoS$_2$ and GaSe films placed inside the microcavity by up to a factor of 60, and directly measure the shortening of the PL life-time in GaSe films by a factor of 10. Both cw and time-resolved observations provide evidence for the Purcell enhancement of the spontaneous emission rate in the studied cavities. 

This work opens a route to a wide range of devices using 2D films as optically active materials and which is also fully compatible with various van der Waals heterostructures. This will lead to realization of cavity-enhanced and tunable light-emitting diodes. An advantage such devices would offer compared with traditional semiconductors is compatibility with a wide range of substrate materials including various dielectrics, polymers and flexible substrates. This would permit a wide choice of materials for fabrication of microcavities (and waveguides), including use of oxides such as SiO$_2$, TiO$_2$ etc and polymers, and also hybrid structures where, for example, oxide dielectric mirrors are combined with polymer cavities. 

Currently, the most advanced metal-chalcogenide heterostructures are built from exfoliated films, and the use of vertical cavity geometry in photonic applications is the most practical and advantageous, as the lateral device sizes are of the order 10-50 $\mu$m only. Use of other geometries such as waveguide, widely employed in semiconductor lasers and modulators, will only be applicable for synthetic films with sizes as large as a few hundreds of micron or several millimeters. 

Finally, we note that excitons in transition-metal dichalcogenides (MoS$_2$, WS$_2$ etc) also exhibit large binding energies and high oscillator strength, and are therefore promising for observation of room temperature polariton effects\citep{LiuArxiv2014}, opening a new field of devices utilizing highly non-linear optical phenomena and cavity QED in a new material system of van der Waals crystals.

\section{Methods}

\subsection{Dielectric mirror fabrication}

An array of concave mirrors used in the top DBR is fabricated using focused ion beam milling of a high flatness silica substrate. The radius of curvature of the concave mirror ranges from 5.6 to 25$\mu$m. The flat DBR (the bottom mirror) in the studied microcavities was fabricated by deposition of 10 pairs of quarter-lambda SiO$_2$/TiO$_2$ layers (with the refractive indexes 1.4 and 2.1) on a high quality silica substrate. For the top DBR, the whole substrate containing the concave pits was coated with 10 pairs of SiO$_2$/TiO$_2$. The DBRs were designed for a center wavelength of 640 nm and provided a stop-band with a width of nearly 200 nm \cite{SI}. 

\subsection{2D film fabrication}

Monolayer MoS$_2$ and thin sheets of GaSe of thicknesses ranging from 30 nm to 100 nm have been obtained by mechanical cleavage of bulk crystals. GaSe films were deposited straight on the flat DBR substrate, whereas the MoS$_2$ films were first deposited on a polymer and then transferred onto the flat DBR using standard transfer techniques \cite{KretininNanoLett2014}. The thickness of the sheets was verified using atomic force microscopy.

\subsection{Optical measurements}

Optical measurements were performed with the DBRs and samples placed in a low pressure He gas at a temperature of 4.2K. Both the flat and concave DBR substrates were attached to closed-loop XYZ nano-positioners with tilt enabling a few tens of nm precision in the translation stage positioning and also large travel range exceeding 1 cm. This permitted two types of PL experiments without warming up the sample to room $T$: (i) PL detection from the film on the flat DBR with the top DBR moved out of the optical path (the 'half-cavity' configuration); (ii) PL from the films placed inside a full cavity, for which the top mirror was moved above the place on the film excited with the laser. For the method (ii) the same location on the film could be studied in several cavities with the top DBRs having different radii of curvature, which was achieved by moving the top DBR in the xy-plane and aligning the desired mirror with the film area excited by the laser. 

A laser diode at 532 nm was used for PL studies in continuous-wave (cw) experiments. For the time-resolved measurements a frequency-doubled pulsed titanium sapphire laser (at 415 nm) was used for excitation and a streak camera was employed for the signal collection. The excitation laser was focused onto the sample with a lens that has a focal length of 7.5 mm (L1 in Fig.\ref{Fig1}), resulting in a spot size on the film of 7 $\mu$m. The size of the spot is negligibly affected by the top DBR. The PL is collected by the same lens and is focused outside the cryostat by another lens (L2 in Fig.\ref{Fig1}) in a fiber delivering light to an 0.5 m spectrometer and a high sensitivity charge coupled device. 

\subsection{Calculations of the PL directionality in a cavity}

The angular width of the fundamental cavity mode was calculated and compared with the radiation pattern for a dipole placed directly on the lower DBR but with no upper mirror. The electromagnetic field associated with the fundamental cavity mode was obtained using the finite difference time domain (FDTD) technique~\cite{meep}. A Fourier transform was applied to the field in the homogeneous region above the structure to obtain the angular power spectrum. The far-field radiation patterns for emitters on the DBR were obtained using a plane-wave expansion of the electric field due to a point dipole source.~\cite{nanooptics,dipoles_interfaces}. Further description of the methods is provided in the supplementary material\cite{SI}. 

Figure~\ref{Fig5} shows the simulated power radiated per unit solid angle in the far field for several mirror radii-of-curvature and separations  as well as for the case without a top mirror. Light emitted upwards only is considered in this plot. In order to simulate the light collection in a finite angle by the objective, the radiation patterns were integrated over the range of polar angles $0\leq \theta \leq \theta_{max}=16.7$ degrees and through the whole range of azimuth angle $\phi$ according to $P = \int^{2\pi}_0\int^{\theta_{max}}_0 p(\theta) sin(\theta) d\theta d\phi$. The angle $\theta$ is defined to be zero in the direction of the center of the objective lens. The value for the power collected by the objective was divided by the power integrated over all directions. 

In the case where no top mirror is present only 9\% of the total power is collected because without a cavity present much of the power is radiated into high angle radiation modes. The cavity modes are much more directional. For cavities C1 and C2 in Fig.~\ref{Fig5} 48\% of total power was radiated in an upwards direction as might be expected since the Bragg mirrors have equal reflectivity. For cavity C3 the beam size at the top mirror inside the cavity is comparable to the mirror radius and scattering at the discontinuity between curved and planar regions becomes important (see Ref.\cite{SI}). Photons scatter out of the confined mode into sideways-propagating modes in the adjacent planar cavity which lowers the ratio of power emitted upwards to 38\%.  

As expected the beam angular spread reduces as the mirror radius of curvature increases. The scattering expected in the case of C3 also affects the angular spread of the beam in the upwards direction. For the cavities C1, C2 and C3, the fractions of power within $\pm$16.7 degrees are 47\%, 44\% and 20\% respectively, giving the 'geometrical' enhancements factors, $F_g$ of 5.3, 4.9 and 2.2 respectively compared to the case with no cavity present. 

\subsection{Calculations of the Purcell enhancement}

Further to the angular distribution, the FDTD method employed above can be used to calculate the effective mode volumes and $Q$-factors for the optical modes in the microcavity. We now discuss how these may be used to make an estimate of the Purcell factor. The enhancement of the spontaneous emission rate of an emitter at position $\vect{r_0}$ due to a cavity may be calculated using the standard formula~\cite{MarkFox}

\begin{equation} \label{F_P}
F_P\left(\vect{r_0}\right) = \frac{3}{4\pi^2}\left(\frac{\lambda}{n\left(\vect{r_0}\right)}\right)^3\left(\frac{Q}{V_{eff}}\right)
\end{equation}

Here $\lambda$ is the vacuum wavelength, $\vect{r_0}$ is the position of the field maximum in the cavity and $n$ is refractive index. The effective mode volume is given by

\begin{equation} \label{eq:veff}
V_{eff} = \frac{\int_{-\infty}^{\infty}\int_{0}^{2\pi}\int_{0}^{\infty}\epsilon(\vect{r})\left|\vect{E}(\vect{r})\right|^2\cdot r\cdot dr d\phi dz}{\epsilon(\vect{r_0})\left|\vect{E}(\vect{r_0})\right|^2}
\end{equation}

Here $\vect{r}=\left(r,\phi,z\right)$ is position in space and $\epsilon(\vect{r})\left|\vect{E}(\vect{r})\right|^2 = \epsilon(\vect{r})\left(\vect{E}(\vect{r})\cdot\vect{E}^\ast(\vect{r})\right)$ is the electric energy density. Since the full electromagnetic fields of the cavity modes are calculated as a function of position by the FDTD simulations the effective mode volumes are obtained simply by numerically performing the integral in equation~\ref{eq:veff}. We obtain volumes of 6.46, 3.10 and 1.61 \micro$m^3$ for the cavities C1, C2 and C3 respectively. 

Our FDTD calculations also give central wavelengths and $Q$-factors for the modes. These are obtained by examining the field in the cavity using a harmonic inversion technique~\cite{meep}. For the cavities C1, C2 and C3 we obtain $Q_1$=11000, $Q_2$=7700 and $Q_3$=4700 respectively which lead to Purcell factors of $F_{P1}$=40.7, $F_{P2}$=59.4 and $F_{P3}$=69.8.

These Purcell factors provide an upper limit to the enhancement which may be achieved in a cavity. The actual ratio of spontaneous emission rates observed in an experiment will depend on the spectral and spatial overlap of the emitter with the optical mode which effect the local density of states and magnitude of the vacuum field respectively~\cite{GerardPRL1998}. Accurate modeling of the experimentally observed cw PL enhancement and lifetime shortening in time-resolved measurements also requires the knowledge of the non-radiative decay rates in the 2D film. \\

{\bf ACKNOWLEDGMENTS \\}

We thank the financial support of FP7 ITN S$^3$NANO, ERC grant EXCIPOL 320570, and the EPSRC Programme Grant EP/J007544/1. A.A.P.T. and J.M.S. acknowledge support from the Leverhulme Trust. D.N.B. and N.N.K. acknowledge support by DPS RAS project II-5-B5.\\

{\bf AUTHOR CONTRIBUTIONS\\}

S.S., S.D., M.S. and F.L. carried out the measurements. F.W. and S.S. fabricated the 2D films. A.A.P.T. and J.M.S. fabricated the concave DBRs. D.N.B. and N.N.K. grew the GaSe crystals. P.M.W. carried out the FDTD calculations. S.S., A.I.T and P.M.W. wrote the manuscript with contributions from M.S., M.S.S., D.N.K., A.A.P.T. and J.M.S. S.S., A.I.T, P.M.W., M.S., M.S.S., D.N.K., S.D., A.A.P.T. and J.M.S. interpreted the results. A.I.T., D.N.K., M.S.S., E.A.C., J.M.S. and K.S.N. advised on various aspects of presented work and managed the project. A.I.T. conceived and led the project.

\bibliography{references1}

\begin{figure*}
\includegraphics[width=0.4\textwidth]{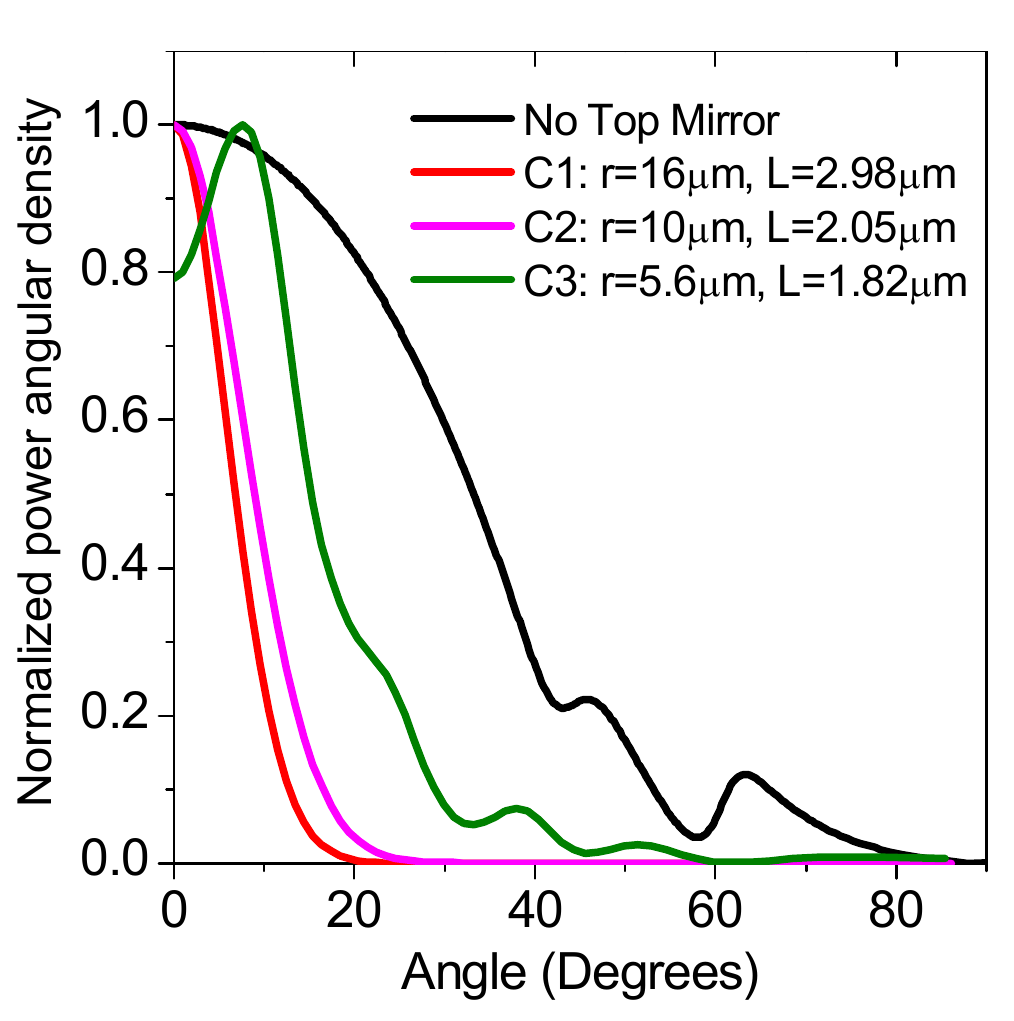}
\caption{\label{Fig5} {\bf Calculated angular distribution of light emission from a 2D film with and without the cavity.} The plot shows the normalized power angular density for the four cases: without the top mirror (the dipole placed on the flat DBR), and for the three different cavities, C1, C2 and C3 with the radii of curvature, $R_c$ and cavity length, $L_{cav}$ given on the graph. In the case where no top mirror is present light emission has a very broad angular spread. The cavity modes are much more directional.}
\end{figure*}

\clearpage

\renewcommand{\thesection}{S\arabic{section}}
\setcounter{section}{0}
\renewcommand{\thefigure}{S\arabic{figure}}
\setcounter{figure}{0}
\renewcommand{\theequation}{S\arabic{equation}}
\setcounter{equation}{0}
\renewcommand{\thetable}{S\arabic{table}}
\setcounter{table}{0}

\renewcommand{\citenumfont}[1]{S#1}
\makeatletter
\renewcommand{\@biblabel}[1]{S#1.}
\makeatother

\pagebreak \pagenumbering{arabic}

\section*{Two-dimensional metal-chalcogenide films in tunable optical microcavities: Supplementary Materials}% Force line 

%\preprint{APS/123-QED}

\section*{}
\textbf{In this Supplementary Materials we present additional details for the sample and device design and fabrication, further characterization data for the microcavities used in the experiments, and, finally, detailed account of the results of FDTD calculations of the optical modes in the external cavity devices used in our work.}

\maketitle
%\thispagestyle{empty}
 %\newpage
%\tableofcontents

%\setcounter{page}{1}
\pagestyle{plain} 
\pagenumbering{arabic}

\section{\label{SI:Samples}Microcavity fabrication and optical properties}

\subsection{Thin film fabrication}
Monolayer MoS$_2$ and thin sheets of GaSe have been obtained by mechanical cleavage of bulk crystals. GaSe films were deposited straight from the wafer dicing tape on the flat DBR substrate, whereas the MoS$_2$ films were first deposited on a polymer layer and then transferred onto the flat DBR using a standard transfer techniques\cite{KretininNanoLett2014}. Fig.\ref{FigS1}(a) shows a microscope image of a thin film of MoS$_2$. Areas with a single- and multiple-monolayer thicknesses can be distinguished on the graph from different shades of green color. The thickness of the films was further verified using atomic force microscopy. As seen on the graph the lateral dimensions of the single-monolayer part of the film exceed 50 $\mu$m by 50 $\mu$m. For GaSe films, flakes with sizes varying from 10 $\mu$m to 50 $\mu$m could be achieved. Our study is focused on relatively thick GaSe films with thicknesses ranging from 30 nm to 100 nm. We find that the PL intensity in GaSe films increases dramatically with the increase of the film thickness.

\subsection{Mirror design and fabrication}
\begin{figure*}
\centering
\includegraphics[width=17 cm]{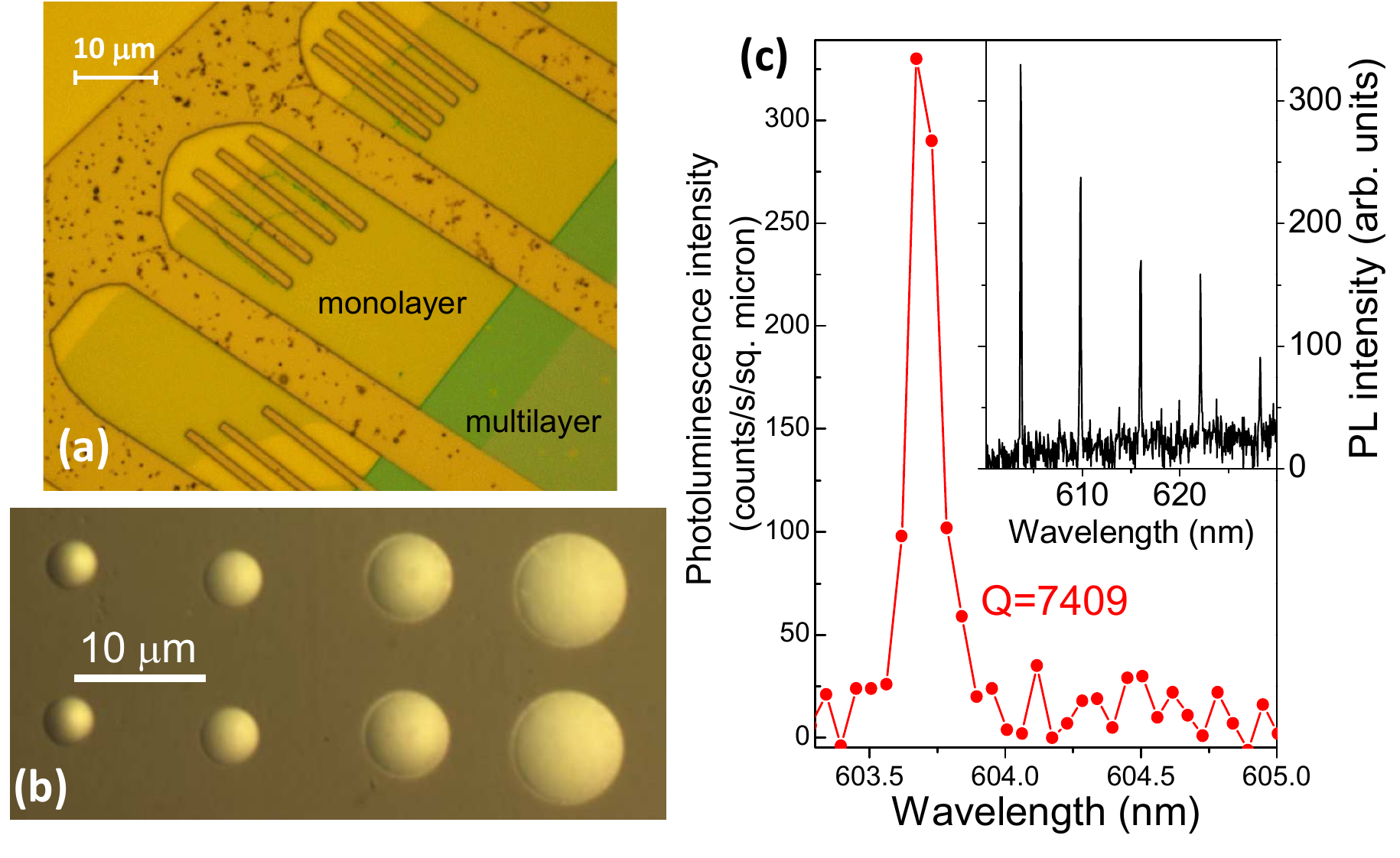}
   \caption{\label{FigS1} (a) Microscope image of a MoS$_2$ thin film deposited on a flat DBR. The monolayer and multilayer parts of the flake are marked. The film is held in position by several gold bars a few tens of nm thick (seen in yellow on the image). (b) Microscope image of a glass plinth with concave mirrors. Concave features with different radii of curvature are milled into a quartz substrate using focused ion beam, and are then covered with ten SiO$_2$/TiO$_2$ quarter-wavelength layer pairs, forming a distributed Bragg reflector. (c) PL spectrum measured in a microcavity containing a thin GaSe film (43 nm). A high $Q$-factor of around 7400 is observed for a longitudinal mode. The inset shows the same PL spectrum in a wide range of wavelength, where other modes with non-zero transverse mode numbers ($m,n\neq$0) are observed.}
\end{figure*}

The open-access microcavity fabrication consists of a two-step process \citep{Dolan2010}. First, the templates of concave mirrors are fabricated using a focused ion beam (FIB) machine (FIB200 from FEI). In this process, gallium ions are fired onto a precisely selected position of a silica substrate for a certain period of time referred to as the 'dwell time'. By adapting the dwell time as a function of the position of the ion beam, we create concave templates with various radii of curvature ranging from 1.7 $\mu$m to 25 $\mu$m onto a single chip. The template dimensions are measured by AFM. The smallest optically active cavities obtained so far had a radius of curvature of 5.6 $\mu$m. The efficiency of the FIB approach relies on the smallest achievable ion beam diameter, which in our case was down to 5 nm.  The rms roughness of the template surface was found to be  below 1 nm. During the second step, both the substrate with the concave mirrors and another flat silica substrate are coated with dielectric distributed Bragg reflectors (DBRs) comprising ten layers of SiO$_2$/TiO$_2$ (with refractive indexes 1.4 and 2.1, respectively) with layer thicknesses tuned to achieve maximum reflectivity at 650 nm. A microscope image of a typical substrate with concave mirrors is shown in Fig.\ref{FigS1}(b).

\subsection{Measurement of high $Q$-factor modes in PL}
Fig.\ref{FigS1}(c) shows photoluminescence (PL) spectra for a cavity comprising a thin GaSe film (43 nm) and having the highest $Q$-factor that we have measured in our devices. As shown in the main panel of the figure, here a longitudinal mode of the cavity has a $Q$-factor of $\approx$7400. This mode has a wavelength $\lambda_{q,m,n}= 603.7$nm. Here $q$=15 and $m,n$=0 (see Eq.1 in the main text for the dependence of the resonant wavelength on $q$, $m$ and $n$). The inset also shows the modes with $q$=14 and $m,n\neq$0 observed in a broader spectral range. The measurements are carried out at a temperature of 4.2K using laser excitation at 532 nm in a cavity with $L_{cav}=$3.25$\mu$m and the radius of curvature of the top mirror $R_c=$25$\mu$m.

\section{\label{SI:Samples}FDTD calculations of the cavity modes}

\subsection{Collection optics considerations}
The ratio of observed intensities for the structure with and without the top mirror depends to some extent on the collection optics since the angular distribution of emission for a given wavelength of light is different for the two cases. The objective lens which collects the light is 7.5mm above the sample. This is sufficiently far from the sample compared to the wavelength of light and to the experimentally measured spot sizes that it can be considered to be in the far field. The collection efficiency is then entirely determined by the lens numerical aperture. The clear aperture of the objective is 4.5mm so only light emitted within $\pm$ 16.7 degrees is collected. To calculate the fraction of light collected it is necessary to know the power radiated per unit solid angle as a function of observation angle in the far field, the so-called radiation patterns~\cite{nanooptics}. Then the power emitted within the collection range of the lens may be compared with the total emitted power. Finally, comparison of the fraction of collected power may be made between the cases with and without top mirror to obtain the collection enhancement due to directionality of the cavity modes.

\subsection{Radiation pattern for electric dipoles on a flat DBR}
We consider first the radiation pattern without the top mirror. The bottom DBR mirror is modelled as a lossless dielectric multilayer consisting of 10 repeats of materials with refractive indexes 1.4 and 2.05 and thicknesses 116.07nm and 79.27nm respectively. The low index material faces the collection lens with the emitters positioned at zero separation from it. The bottom of the DBR rests on a semi-infinite glass substrate with refractive index 1.54. The coordinate axes are chosen so that the layer planes are perpendicular to the z-axis. The emission from the sample may be modelled by an incoherent ensemble of classical electric point dipole current sources. Since in this work the excitons are two-dimensional it is assumed that the emitters are oriented randomly in the plane of the monolayer material. To model the random distribution of in-plane polarisations one may take the incoherent sum of the radiation patterns due to any two orthogonally polarised in-plane dipoles. The resulting total radiation pattern must then have circular symmetry about the $z$ axis since there is no preferred azimuthal direction. As the spot size is small it is sufficient to treat only dipoles positioned at the origin $r=0$ as the radial position will have little effect on the angular spread of the emission. Calculations are performed for dipoles radiating at a frequency corresponding to a wavelength of 680 nm in free space, as in the experiment. We follow the method presented in references~\cite{nanooptics,dipoles_interfaces}. The electric field due to a point electric dipole current radiating in free space may be expanded in a basis of plane-waves. In the presence of the mirror the total upwards-radiating field (towards the collection lens) is the coherent sum of the radiation field emitted upwards by the dipole and the reflection from the mirror of the field emitted downwards by the dipole. The free-space fields are separated into components with TE and TM polarisation with respect to the planar multilayer and the amplitude reflection coefficients are calculated using a transfer matrix technique.

\begin{figure}
\centering
\includegraphics[width=8.5 cm]{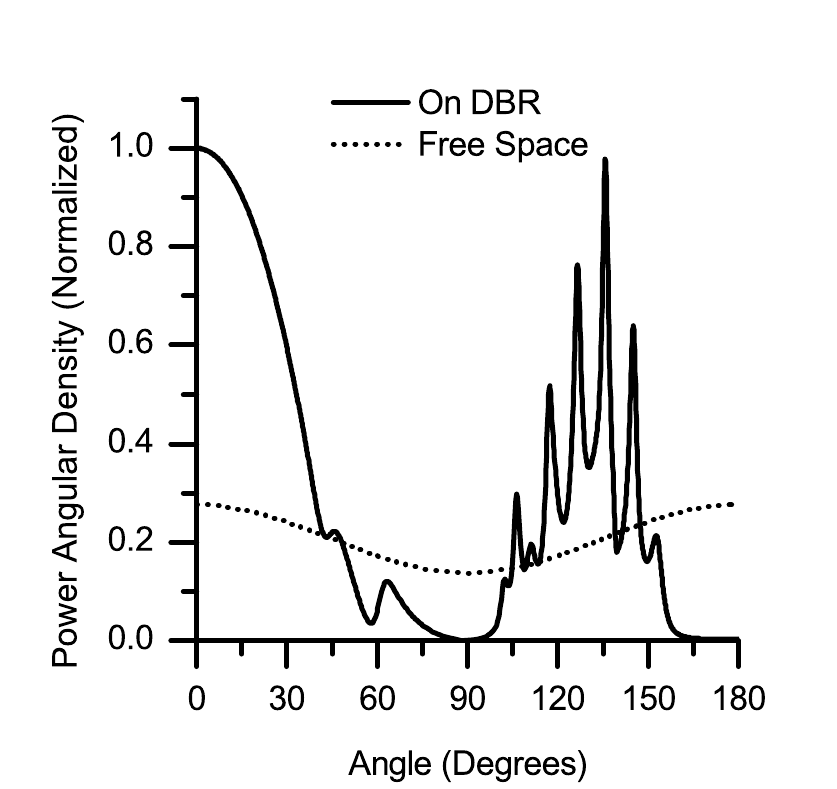}
\caption{Dependence of emitted power per unit solid angle on observation angle in air for randomly oriented dipoles in the x-y plane on the DBR and in free space. Both curves have been normalised to the peak value for dipoles on the DBR.}
\label{fig:dipole_DBR}
\end{figure}

Figure~\ref{fig:dipole_DBR} shows the sum of radiation patterns due to $x$ and $y$ polarised dipoles for the case of dipoles in free space and directly on top of the DBR. For zero angle the DBR reflective phase is close to zero so the reflection reinforces the upward propagating wave and the upwards emission is enhanced with respect to free space. At larger angles the reflective phase increases so that the reflection begins to interfere destructively with the directly radiated field. For angles greater than 90 degrees the radiation is into the substrate. Very little power is radiated into modes close to 180 degrees because the DBR reflects them. For angles greater than 20 degrees from the negative $z$-axis, however, the DBR becomes ineffective and light is lost into the substrate. To obtain total power radiated in the range of polar angles $0\leq \theta \leq \theta_{max}$ the radiation patterns must be integrated with azimuthal and polar collection angles according to $P(\theta_{max}) = \int^{2\pi}_0\int^{\theta_{max}}_0 p(\theta,\phi) sin(\theta) d\theta d\phi$ where $sin(\theta) d\theta d\phi$ is the differential element of solid angle. Overall, 49\% of the total radiation is emitted in the upwards direction and 9\% is emitted within $\pm$16.7 degrees. This compares to only 3\% emitted within $\pm$16.7 degrees for dipoles in free space.  

\begin{figure*}
\centering
\includegraphics[width=17 cm]{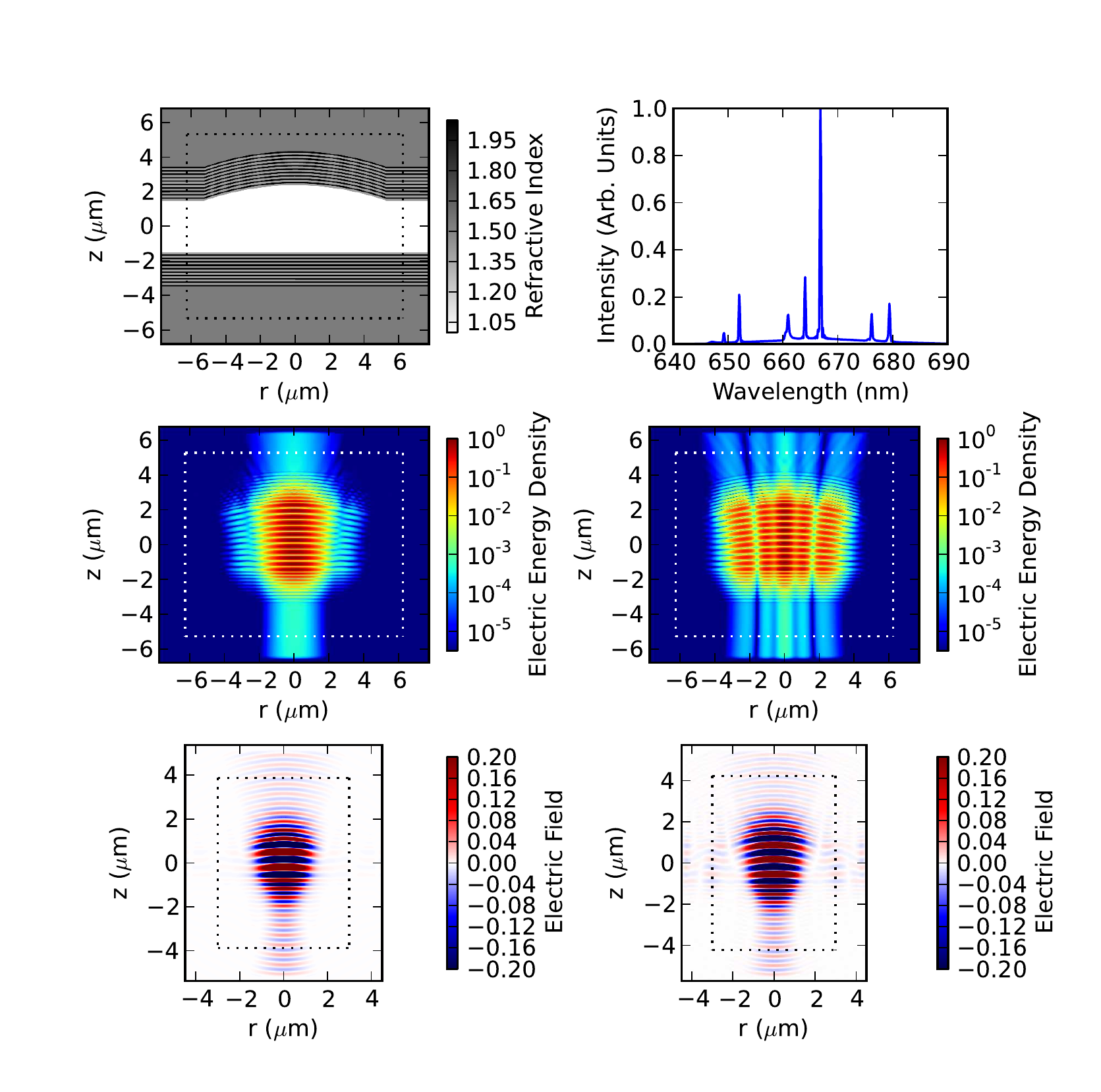}
\caption{(a) Dielectric profile of cavity with 16\micron radius of curvature mirror used in FDTD simulation. The dotted lines show the position of flux planes used to monitor the energy flux out of the structure. (b) Emission spectrum from the structure in (a) after excitation with a broadband source. (c,d) Electric density $E\cdot E^{\ast}$ for two modes in (b) on a logarithmic colour scale. (e,f) Radial electric field component on an exaggerated colour scale for 5.6\micron radius of curvature cavities with mirror separations 1.12 and 1.82 respectively.}
\label{fig:FDTD}
\end{figure*}

\subsection{Modes in a cavity with a concave DBR}
When the top mirror is present a cavity is formed which confines electromagnetic modes at certain frequencies in all three dimensions. At these resonant frequencies the fields and hence the angular spread of the emission are determined by the cavity geometry. Simulations of the electromagnetic fields associated with the resonant cavity modes were performed with the finite-difference time-domain (FDTD) method, using a freely available software package~\cite{meep}. The fields were first determined in the near-field zone close to the cavity. Simulations were performed in a cyclindrical geometry on a two-dimensional grid of radial $r$ and axial position $z$. This allowed much faster simulation times than a full three-dimensional calculation, which was necessary due to the rather small 10 nm grid resolution used to accurately represent the cavity layers and top mirror curvature. 

Figure~\ref{fig:FDTD}(a) shows a schematic of the dielectric profile used for the simulations. The lower DBR is the same as in the planar case discussed earlier. The upper DBR is of the same structure as the lower DBR with low index material adjacent to the cavity. In the radial range $0 \leq r \leq r_m$, where $r_m$ is the mirror radius, the DBR structure was offset in the positive $z$-direction according to $z = z_0 + \sqrt{R_c^2 - r^2} - \sqrt{R_c^2 - r^2_m}$ where $R_c$ is the radius of curvature.

The separation between the upper and lower mirrors was first set to the experimentally estimated value and then refined in the following way. At the start of the simulation a broad frequency spectrum of electromagnetic radiation was excited by an electric dipole current source with short Gaussian temporal profile positioned on top of the lower DBR at $r=400$nm. The electromagnetic energy flux passing through a box surrounding the structure (denoted by dotted lines in Fig.~\ref{fig:FDTD}(a)) was collected for a sufficiently long time to allow all the energy to leave the simulation region. The flux was Fourier transformed to obtain a spectrum of the radiation emitted by the structure. Such a spectrum is shown in figure~\ref{fig:FDTD}(b). The sharp peaks identify resonant cavity modes and correspond to modes with a range of longitudinal (z-direction) and transverse (radial) quantization numbers. 

To find the fundamental (zero transverse quantization number) mode the simulation was repeated a number of times using a narrow-band excitation centered on each cavity mode in the spectrum. The cavity modes have a high quality factor $Q$ and so decay much more slowly than other transient fields caused by the excitation. After several decay-times of the chosen cavity mode the remaining electric and magnetic fields may be considered to have an approximately single-frequency harmonic time dependence and to represent the spatial dependence of the chosen cavity mode. Simulations were run for between one and three decay times $\tau = Q/\omega$ in order to reach this condition. After this the electric and magnetic fields at all points in the simulation volume were output. Figures~\ref{fig:FDTD}(c-d) show the spatial profiles of the time-averaged electric intensity $E \cdot E^{\ast}$ for two of the modes in the spectrum corresponding to the longitudinal mode (c) and a mode with non-zero radial quantization number (d). The mirror separation was then adjusted slightly to bring the fundamental mode close to the experimental wavelength before the fundamental mode field profile was recalculated.

Figures~\ref{fig:FDTD}(e-f) show the radial component of electric field as a function of position for two different mirror separations and 5.6\micron radii of curvature. As can be seen from Fig.~\ref{fig:FDTD}(c,e) the upwards radiation forms a beam centered about $r=0$ which falls off to negligible intensity by the edge of the top flux plane. Comparing the fields above and below the structure we see that the curvature of the top mirror has a lensing effect which makes the beam propagating from the top of the sample spread more compared to that from the bottom. To obtain the relevant portion of the radiation pattern the total energy flux through the four flux planes was first examined and the fraction passing through the top compared to the total obtained. The field above the structure was then Fourier transformed to determine the spread of upwards power among waves propagating with different in-plane wavevectors and hence at different angles. Finally, these were integrated in the same manner discussed above for the case without top mirror. The input parameters and simulation results are summarised in table~\ref{table:sim_params}.

\begin{table}
\caption{\label{table:sim_params}Simulation Parameters and Results}
	\begin{ruledtabular}
	\begin{tabular}{l c c c c c}
		\parbox{1.0 cm}{Cavity}	&	\parbox{1.4 cm}{Radius of curvature (\micro m)}	&	\parbox{1.6 cm}{Mirror Separation	(\micro m)} &	\parbox{1.7 cm}{Mode Wavelength (nm)} & \parbox{1.1 cm}{Quality Factor}	& \parbox{1.5 cm}{Power Collected (\%)}	\\ \hline
		C1			&	16							&	2.98							&	679.8						& 11000			& 47										\\
		C2			& 10							& 2.05							& 681.1						&	7700			& 44										\\
		C3			&	5.6							&	1.82							&	678.9						&	4700			& 20										\\
		C4			&	5.6							&	1.12							&	678.1						& 5500			& 31										\\	
		C5			& 5.6							& 2.52							& 679.9						& 1900			& 6											\\
	\end{tabular}
	\end{ruledtabular}
\end{table}

\subsection{Effect of the concave DBR on the radiation pattern}
\begin{figure}[b]
\centering
\includegraphics[width=8 cm]{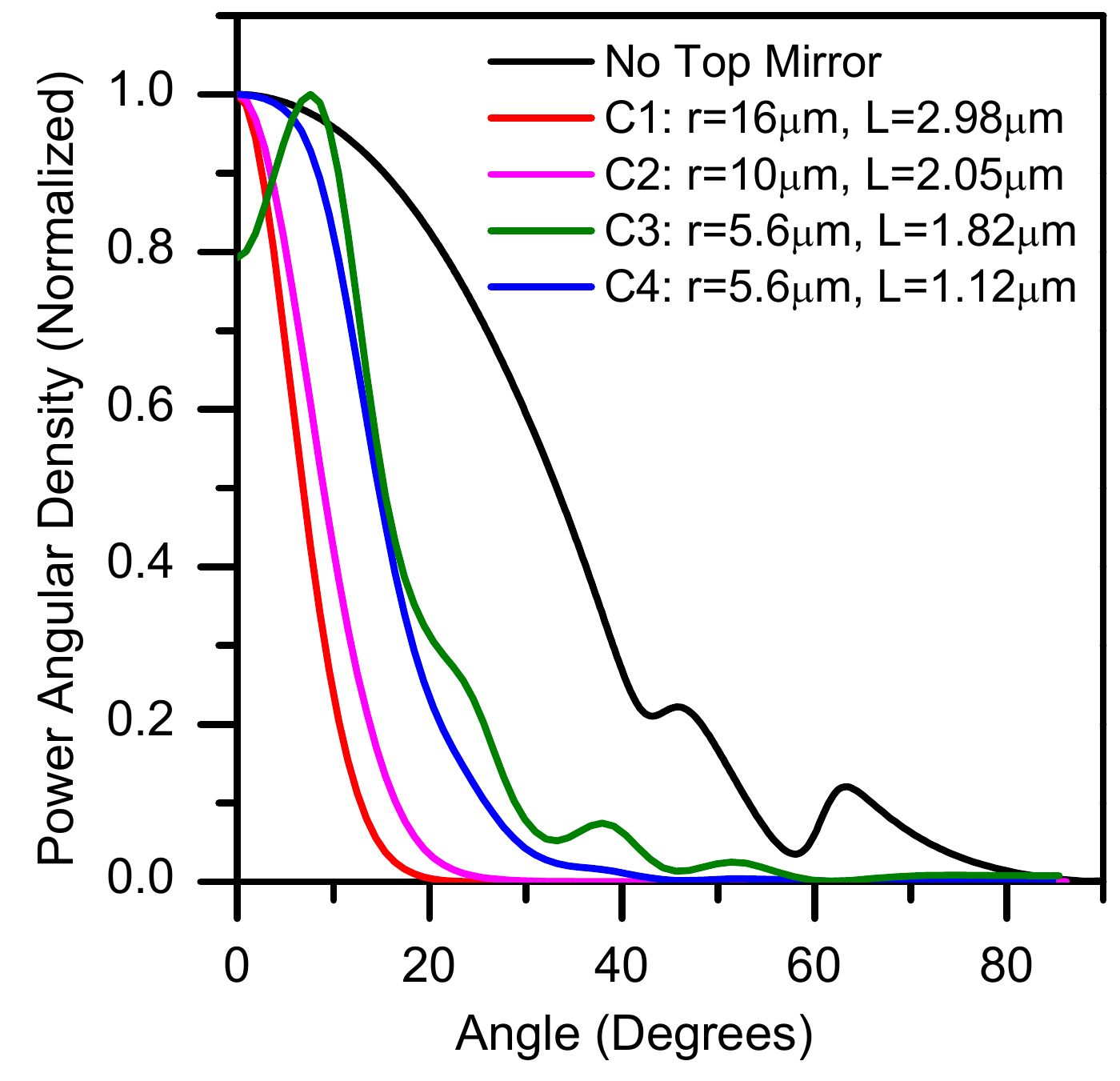}
\caption{Dependence of emitted power per unit solid angle on observation angle in air for several cavities and without top mirror.}
\label{fig:angular}
\end{figure}

Angular profiles for radiation from several cavities from Table~\ref{table:sim_params} are presented in Fig.~\ref{fig:angular}. We observe significant narrowing of the angular distribution for emitters in the cavities with concave mirrors compared to the case of a flat DBR only. The observed dependences for different cavities also agree with the qualitative notion that the larger cavity should give a more directional beam. Here we would like to discuss in more detail behavior of the cavity with the 5.6 $\mu$m radius mirror, as new features are observed in the radiation pattern for some mirror separations in this configuration.

Figures~\ref{fig:FDTD}(e-f) correspond to the angular profiles presented in Fig.~\ref{fig:angular} for 5.6\micron radius of curvature and 1.1\micron and 1.8\micron mirror separations. For the larger mirror separation we see an increase in the field amplitude in the planar regions adjacent to the curved mirror and simultaneously the appearance of extra non-zero wavevector components in the angular spectrum of upwards propagating radiation. These are accompanied by a drop in cavity quality factor from 5500 to 4700 and an increase in the proportion of energy flowing through the side flux planes. The quality factor and sideways energy loss get worse with further increases in mirror separation, see Table~\ref{table:sim_params} structure C5. It has been shown experimentally that the quality-factor in hemispherical cavities increases with increasing mirror separation up to a critical value where it begins to decrease~\cite{DufferwielAPL2014}. This behaviour was attributed to a loss of mode stability which, in a purely geometrical picture, is where some rays at higher angles become able to exit the resonator on each round trip. Geometrical arguments predict that this occurs when the mirror separation is greater than the radius of curvature~\cite{DufferwielAPL2014}. However, these arguments ignore the finite diameter of the curved section of the mirror. In the real system there is a sharply discontinuous interface between curved and planar regions which may cause scattering with a strength which depends on the local field amplitude. Qualitatively, the angular spread of the mode is dictated by the radius of curvature. The spatial extent of the beam at the top mirror will be proportional to this angular spread and the mirror separation. When the separation is large the spatial beam size will overlap the discontinuous region leading to scattering. It is likely that this scattering is the cause of the observed energy loss from the confined mode into sideways propagating modes and also of the complicated angular emission profile above the structure. The two effects together tend to reduce the fraction of power collected by the objective lens.

Finally, the fractions of total emitted power collected by the objective may be compared between the cavities and the case where there is no mirror (see main text and also Methods). For the cavities close to the experimental parameters C1 and C3 we expect to collect 5.3 and 2.2 times more of the total emission than in the case with no top mirror. This agrees with the qualitative notion that the larger cavity should give a more directional beam and so more of the emission should be collected.

\clearpage
\bibliography{references1}

%\subsection{A Subsection}

\end{document}